\documentclass{epl}%
\usepackage{amssymb}
\usepackage{psfig}
\usepackage{amsmath}
\usepackage{amsfonts}
\usepackage{graphicx}%
\begin{document}

\title{The Casimir force between rough metallic plates}
\author{C. Genet \inst{1} 
\thanks{Present address: Huygens Laboratory - Universiteit
Leiden, P.O. Box 9504, 2300 RA Leiden, The Netherlands}
\and A. Lambrecht\inst{1}
\and P. Maia Neto\inst{2}
\and S. Reynaud\inst{1} 
\thanks{E-mail: \email{reynaud@spectro.jussieu.fr} ; 
Url: \email{www.spectro.jussieu.fr/Vacuum} } }
\shortauthor{C. Genet \etal}
\institute{ \inst{1} Laboratoire Kastler
Brossel \thanks{Laboratoire du CNRS, de l'\'Ecole Normale
Sup\'erieure et de l'Universit\'e Pierre et Marie Curie.} -
case 74, Campus Jussieu, 75252 Paris, France\\
\inst{2} Instituto de F\'{\i}sica - UFRJ, caixa postal 68528,
21945-970 Rio de Janeiro, Brazil}

\recff{Feb. 7, 2003}{March 19, 2003}

\maketitle

\begin{abstract}
The Casimir force between two metallic plates is affected by their roughness
state. This effect is usually calculated through the so-called `proximity
force approximation' which is only valid for small enough wavevectors in the
spectrum of the roughness profile. We introduce here a more general
description with a wavevector-dependent roughness sensitivity of the Casimir
effect. Since the proximity force approximation underestimates the effect, a
measurement of the roughness spectrum is needed to achieve the desired level
of accuracy in the theory-experiment comparison.

\end{abstract}

\section{Introduction}

The Casimir force \cite{Casimir48} has recently been measured with a good
experimental precision which allows for an accurate comparison between
measured values and theoretical predictions \cite{Bordag01,Lambrecht02}. This
comparison plays an important role in the searches of new weak forces with
nanometric to millimetric ranges motivated by theoretical unification models
\cite{Bordag99,Long99,Fischbach01}. These forces would appear as
experiment/theory differences in precise measurements of the Casimir force. As
far as an accurate theory-experiment comparison is aimed at, the accuracy of
theory is as crucial as the precision of experiments. If the target is a given
accuracy, say at the $1\%$ level, the theoretical prediction has to be
mastered at this level as well as the experimental measurement. Clearly, this
requires to take into account the real conditions of the experiments which
differ from the ideal situation often assumed in the theory of the Casimir effect.

Casimir initially studied the case of a pair of perfectly smooth, flat and
parallel plates in the limit of zero temperature and perfect reflection. He
found a force which depends only on geometrical properties, the distance $L$
between the plates and their area $A$ supposed to be much larger than $L^{2}$,
and two fundamental constants, $c$ and $\hbar$. This remarkable fact is
related to the assumption of perfect reflection while the most precise
experiments \cite{Harris00,Chan01} are performed with metallic mirrors which
are nearly perfect reflectors only at frequencies smaller than a
characteristic plasma frequency $\omega_{P}$ so that the force is
affected by imperfect reflection at distances smaller than or of the order of
the plasma wavelength $\lambda_{P}=\frac{2\pi c}{\omega_{P}}$. 
The same experiments are performed at room temperature, which implies that
the force also depends on the scattering of blackbody radiation, in a geometry
different from the ideal geometry considered by Casimir: the force is measured
between a plane and a spherical reflectors which show some surface roughness.

The aim of the present letter is to raise questions about the validity of the
current evaluation of the effect of surface roughness. For the sake of
comparison with forthcoming discussions, we first recall the principles of
this evaluation which is based on the proximity force approximation
\cite{Deriagin68}, denoted PFA hereafter. We then introduce an improved method
where the sensitivity to roughness depends on the wavevector associated with
surface deformations. We get information on this sensitivity by considering
metallic plates in the two limits of long or short distances,
where the roughness sensitivity can be deduced from earlier calculations
devoted respectively to perfect reflectors \cite{Emig01} and to mirrors
described by the surface plasmon approach \cite{Maradudin80,Mazur81}. In the
present letter, we focus our attention on questions of interest for
the comparison of theoretical predictions and experimental measurements of the
Casimir force in the plane-sphere geometry. We disregard the
temperature effect which is significant at large distances whereas roughness
corrections are more important at the smallest distances explored in the
experiments. We give only the main results of our calculations and
refer the reader interested by more detailed developments to forthcoming publications.

\section{The proximity force approximation}

When using the proximity force approximation (PFA), the force $F_{\mathrm{PS}}$ 
in the plane-sphere geometry is obtained as the sum of the contributions corresponding 
to an effective inter-plate distance $z$ which varies from the distance of closest
approach $L$ to infinity. As the area element $\mathrm{d}A=2\pi R\mathrm{d}z$
corresponding to an interval $\mathrm{d}z$ is proportional to this interval
and to the radius $R$ of the sphere, $F_{\mathrm{PS}}$ is given by geometric
arguments and by the Casimir energy $E_{\mathrm{PP}}$ calculated between two
planes
\begin{equation}
F_{\mathrm{PS}}\left(  L\right)  =2\pi R\frac{E_{\mathrm{PP}}\left(  L\right)
}{A}\quad,\quad E_{\mathrm{PP}}\left(  L\right)  =\int\limits_{L}^{\infty
}\mathrm{d}z\ F_{\mathrm{PP}}\left(  z\right)  \label{FPSsmooth}
\end{equation}
We emphasize that the PFA amounts to the addition of contributions
corresponding to different distances $z$, assuming these contributions to be
independent. But the Casimir force is not additive, so that the PFA cannot be
exact, although it is often improperly called a theorem. The results available
for the plane-sphere geometry \cite{Langbein71,Keifer78,Balian78} show that
the PFA leads to correct results when the radius $R$ of the sphere is much
larger than the distance $L$ of closest approach.

The roughness correction is also often evaluated by using the PFA. We suppose
that the two mirrors $\left(  i=1,2\right)$ have roughness profile functions
$h_{i}\left(  \mathbf{r}\right) $ where $\mathbf{r}$ collects the two
transverse coordinates $\left( x,y\right)$ orthogonal to the direction $z$
of the cavity. Both deformations $h_{1}\left( \mathbf{r}\right) $ and
$h_{2}\left( \mathbf{r}\right) $ are counted as positive when they
correspond to length increases above the mean. Using calligraphic letters for
the energy between rough plates and normal letters for the energy between
smooth plates, we obtain with the PFA
\begin{equation}
\mathcal{E}_{\mathrm{PP}}\left(  L\right) =\left\langle E_{\mathrm{PP}}
\left(  L+h\right)  \right\rangle \simeq E_{\mathrm{PP}}\left( L\right)
+\frac{E_{\mathrm{PP}}^{\prime\prime}\left( L\right) }{2}\left\langle
h^{2}\right\rangle \quad,\quad h=h_{1}+h_{2} 
\label{pfa}
\end{equation}
The symbol $\left\langle \ldots\right\rangle $ denotes an average over the
transverse coordinates. We have assumed that the profiles have a null average
value $\left\langle h_{1,2}\right\rangle =0$ and restricted the evaluation at
the second order in the deformations. Hence, the roughness correction depends
on the second order derivative $E_{\mathrm{PP}}^{\prime\prime}\left(
L\right)$ of the energy and on the variance of the length deformation $h$.
This expression is equivalent to the procedure used for analyzing the effect
of roughness in recent experiments \cite{Harris00,Klimchitskaya99}.

It can be simplified one step further when the area $A$ of the plates contains
a large number of correlation areas. We denote $\ell_{C}$ the correlation 
length of the roughness profiles and suppose $A\gg\pi \ell_{C}^{2}$. 
We also suppose that the profiles of the two plates
have no special relation to each other so that the integration over the
surface is equivalent to a statistical averaging over a number of different
realisations of these profiles. It follows that the cross correlation
$\left\langle h_{1}h_{2}\right\rangle $ between the two profiles can be
ignored, which leads to the expression
\begin{equation}
\mathcal{E}_{\mathrm{PP}}\left(  L\right)  \simeq E_{\mathrm{PP}}\left(
L\right)  +\frac{E_{\mathrm{PP}}^{\prime\prime}\left(  L\right)  }{2}
a^{2}\quad,\quad a^{2}=\left\langle h_{1}^{2}\right\rangle +\left\langle
h_{2}^{2}\right\rangle \quad,\quad A\gg\pi\ell_{C}^{2}
\label{EPPpfa}
\end{equation}
Note that the last simplification would not be applicable to the case of
corrugated plates \cite{Chen02} which is not considered in the present letter.

The preceding equation can be translated into expressions of the
Casimir forces between rough plates, denoted with similar conventions. In the
plane-plane geometry, the force $\mathcal{F}_{\mathrm{PP}}$ between rough
plates would be obtained by differentiating (\ref{EPPpfa}) with respect to
$L$. Here we consider the plane-sphere geometry with the radius of the sphere
large enough so that the PFA (\ref{FPSsmooth}) is valid in the absence of
roughness. In the presence of roughness, we 
define $R$ as the radius of the sphere realizing the best
fit of the real surface of the curved mirror. We then define the
roughness profile as the deviation of the real surface from the best-fit
sphere and suppose the statistical properties of this profile to be
uncorrelated with the mean spherical geometry. We also assume that the number
of correlation areas $\pi\ell_{C}^{2}$ contributing significantly to
the whole force is very large, so that the integration over the surface is
still equivalent to a statistical averaging over a number of different
realisations of these profiles. This assumption is consistent when the
condition $\pi RL\gg\pi\ell_{C}^{2}$ holds (see a similar discussion
in \cite{Klimchitskaya99}) and it allows us to ignore the cross correlation
between the profiles of the two mirrors. When these conditions are
met, the force $\mathcal{F}_{\mathrm{PS}}$ is given by an expression which
generalizes (\ref{FPSsmooth}) to the case of rough plates
\begin{equation}
\mathcal{F}_{\mathrm{PS}}\left(  L\right)  =2\pi R\frac{\mathcal{E}%
_{\mathrm{PP}}\left(  L\right)  }{A}\quad,\quad R\gg L\quad,\quad RL\gg
\ell_{C}^{2} 
\label{FPSexp}
\end{equation}

\section{The spectral sensitivity to roughness}

Clearly, the PFA can only be valid for long-wavelength deformations. In
particular, it gives a correct evaluation of the effect of curvature of the
spherical mirror when $R\gg L$. As far as roughness is concerned, 
the PFA holds for small enough values of the transverse
wavevector $\mathbf{k}=\left(  k_{x},k_{y}\right)  $ but not in the general
case of an arbitrary wavevector.

In order to deal quantitatively with this problem, we introduce the spectral
densities $\sigma_{ij}$ which describe the correlation properties of the
profiles $h_{i}$
\begin{equation}
\sigma_{ij}\left[  \mathbf{k}\right]  
= \int \mathrm{d}^{2}\mathbf{r} \ \ e^{-i\mathbf{k.r}}
\left\langle h_{i}\left( \mathbf{r}\right) h_{j}\left( \mathbf{0}\right)
\right\rangle 
\end{equation}
We write the Casimir effect at the second order in 
roughness profiles as
\begin{equation}
\mathcal{E}_{\mathrm{PP}}\left(  L\right)  \simeq E_{\mathrm{PP}}\left(
L\right)  +\sum_{i,j}\int\frac{\mathrm{d}^{2}\mathbf{k}}{4\pi^{2}}
\ G_{ij}\left[  \mathbf{k}\right]  ~\sigma_{ij}\left[  \mathbf{k}\right]
\end{equation}
The functions $G_{ij}$ reduce to $E_\mathrm{PP}^{\prime\prime}/2$ 
in the PFA so that the preceding
expression reduces to (\ref{pfa}). In the general case, the functions $G_{ij}$
depend on the transverse wavevector $\mathbf{k}$ and thus measure the spectral
sensitivity to roughness of the Casimir effect. As previously, we disregard
the crossed correlation terms $\sigma_{ij}$ with $i\neq j$ which tend to be
averaged to zero in the integration over the surface. For the sake of
simplicity, we also suppose that the two mirrors are made with the same metal
so that the roughness sensitivity is described by a single function
\begin{align}
&  \mathcal{E}_{\mathrm{PP}}\left(  L\right)  \simeq E_{\mathrm{PP}}\left(
L\right)  +\int\frac{\mathrm{d}^{2}\mathbf{k}}{4\pi^{2}}\ G\left[
\mathbf{k}\right]  ~\sigma\left[  \mathbf{k}\right]  \quad,\quad\sigma\left[
\mathbf{k}\right]  =\sigma_{11}\left[  \mathbf{k}\right]  +\sigma_{22}\left[
\mathbf{k}\right]  \nonumber\\
&  G\left[  \mathbf{k}\right]  =G_{11}\left[  \mathbf{k}\right]
=G_{22}\left[  \mathbf{k}\right]  =\frac{E_{\mathrm{PP}}^{\prime\prime}\left(
L\right)  }{2}\rho\left[  \mathbf{k}\right]  \label{EPP}%
\end{align}
We have introduced a reduced sensitivity $\rho$ which goes to
unity in the PFA sector. Due to the cylindrical symmetry with respect to
rotations in the transverse plane, $G$ and $\rho$ are functions
of $\left\vert \mathbf{k}\right\vert $ only.

The derivation of equation (\ref{FPSexp}) presented above remains valid outside 
the PFA sector. Hence, the roughness correction to the Casimir force in
the plane-sphere geometry is read
\begin{equation}
\frac{\mathcal{F}_{\mathrm{PS}}\left(  L\right)  -F_{\mathrm{PS}}\left(
L\right)  }{F_{\mathrm{PS}}\left(  L\right)  }=\frac{E_{\mathrm{PP}}
^{\prime\prime}\left(  L\right)  }{2E_{\mathrm{PP}}\left(  L\right) }
\int\frac{\mathrm{d}^{2}\mathbf{k}}{4\pi^{2}}\ \rho\left[  \mathbf{k}\right]
\ \sigma\left[  \mathbf{k}\right]  \label{FPS}%
\end{equation}
Incidentally, we notice that the Casimir force $\mathcal{F}_{\mathrm{PP}}$ in
the plane-plane geometry could be obtained by differentiating $\mathcal{E}%
_{\mathrm{PP}}$ with respect to $L$, taking into account that $\rho$ depends
on the distance $L$. For the sake of illustrating the significance of these
results, we can consider the example of a Gaussian roughness spectrum
\begin{equation}
\sigma\lbrack\mathbf{k}]=a^{2}\pi\ell_{C}^{2}\exp\left(
-\frac{\mathbf{k}^{2}\ell_{C}^{2}}{4}\right)  
\label{Gaussian}
\end{equation}
where $a$ measures the roughness amplitude and $\ell_{C}$ the
roughness correlation length. This example allows one to specify the
definition of the correlation length $\ell_{C}$ but it cannot be
used as a general description of roughness profiles. In order to discuss
the experimental situations, the roughness spectra must be deduced 
from images of the real plates.

\section{Metallic mirrors in the limit of long distances}

\begin{figure}[ptb]
\centerline{\psfig{figure=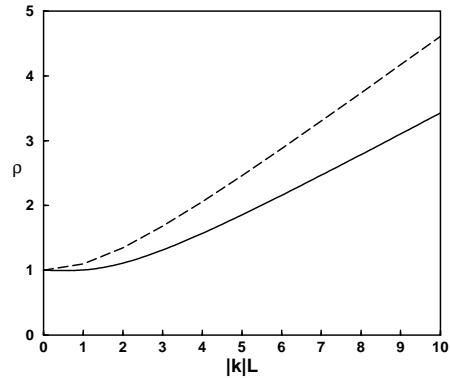,width=7cm}}\caption{Variation of $\rho$
versus $\left\vert \mathbf{k}\right\vert L$ in the two regimes of long
distances ($L \gg\lambda_{P}$, solid line) and short distances ($L
\ll\lambda_{P}$, dashed line).}
\label{Rho}
\end{figure}

When the inter-plate distance $L$ is much larger than the plasma length
$\lambda_{P}$, metallic mirrors behave as perfect reflectors. It follows
that the roughness sensitivity can be obtained as a by-product of recent
computations by Emig \textit{et al} of the Casimir force between a corrugated
mirror and a flat one, both being perfect reflectors
\cite{Emig01}. Emig \textit{et al} have evaluated the effect on the Casimir
force of a periodic perturbation with a wavelength $\lambda$ of one of the two
plates. Equation (4) of \cite{Emig01} can be interpreted as giving the 
function $\rho$ at the wavevector $\left\vert \mathbf{k}\right\vert
=2\pi/\lambda$
\begin{equation}
\rho\left[  \mathbf{k}\right]  =\frac{G_{\mathrm{TM}}\left(  s\right)
+G_{\mathrm{TE}}\left(  s\right)  }{G_{\mathrm{TM}}\left(  0\right)
+G_{\mathrm{TE}}\left(  0\right)  }\quad,\quad s\equiv\frac{L}{\lambda}
=\frac{\left\vert \mathbf{k}\right\vert L}{2\pi} 
\label{RhoLong}
\end{equation}
The functions $G_{\mathrm{TM}}$ and $G_{\mathrm{TE}}$ correspond to the
contributions of TM and TE modes and are given by equations (5,6) of
\cite{Emig01}. They reduce to the PFA limit when $s\rightarrow0$, a property
here included in the very definition of $\rho\left[  \mathbf{k}\right]  $. The
variation of $\rho$ versus the dimensionless factor $\left\vert \mathbf{k}
\right\vert L$ is shown as the solid line on Figure \ref{Rho}.

The PFA limit $\rho\simeq1$ is recovered for $\left\vert
\mathbf{k}\right\vert L$ smaller than or of the order of unity. This means
that the PFA (\ref{EPPpfa}) leads to reliable evaluations 
when the wavevectors which contribute significantly to
surface roughness lie in the PFA sector $\left\vert \mathbf{k}\right\vert
L\lesssim1 $. When this is not the case, the factor $\rho$ is almost
everywhere larger than unity, which means that the effect of roughness 
is underestimated in the PFA analysis. In the limiting case of large
wavevectors, $\rho$ is found to reach large values proportional to 
$\left\vert \mathbf{k}\right\vert $ \cite{Emig01} 
\begin{equation}
\rho\left[  \mathbf{k}\right]  \simeq\beta\left\vert \mathbf{k}\right\vert
L\quad,\quad\beta=\frac{1}{3}\quad,\quad\left\vert \mathbf{k}\right\vert
L\gtrsim1 
\label{LimLong}
\end{equation}
This means that the error in the PFA
estimation of the roughness correction may be important.

\section{Metallic mirrors in the limit of short distances}

We now consider the limit $L\ll\lambda_{P}$ of short distances, where
the Casimir energy can be written in terms of the dispersion relation
characterizing the surface plasmons \cite{vanKampen68,Barton79}. The change of
the dispersion relation and of the Casimir energy by the surface roughness has
been studied by Maradudin and Mazur \cite{Maradudin80,Mazur81}. The roughness
sensitivity can be deduced from these results through a chain of
reinterpretations and substitutions.

Equation (25) of \cite{Mazur81} can be rewritten as giving the second order
correction to the Casimir energy between two rough planes
\begin{equation}
\mathcal{E}_{\mathrm{PP}}\left(  L\right)  \simeq E_{\mathrm{PP}}\left(
L\right)  -\hbar A\int\frac{\mathrm{d}^{2}\mathbf{q}}{4\pi^{2}}\frac
{\mathrm{d}^{2}\mathbf{p}}{4\pi^{2}}~\sigma\left[  \mathbf{q-p}\right]
\int\limits_{0}^{\infty}\frac{\mathrm{d}\xi}{2\pi}T\left[  \mathbf{q}
,\mathbf{p},i\xi\right]  
\label{EPPplasmon}
\end{equation}
This is the contribution of TM modes, with TE modes having a negligible
contribution at short distances. As previously, we consider the case of 
uncorrelated profiles on the two plates and deduce the kernel
$T$ from the $T_{2}$ of equations (26c,28) in \cite{Mazur81}
through the substraction $T=T_{2}\left[  L\right]  -T_{2}\left[
L\rightarrow\infty\right]  $. We write this kernel in terms of the reflexion
amplitudes $r_{1}=r_{2}=r$ for TM modes near grazing incidence, assuming the
mirrors to be made with the same metal described by the plasma model. We obtain
after a number of manipulations
\begin{align}
T\left[  \mathbf{q},\mathbf{p},i\xi\right]   &  =\frac{\left\vert
\mathbf{q}\right\vert }{1-r^{2}e^{-2\left\vert \mathbf{q}\right\vert L}}
\frac{\left\vert \mathbf{p}\right\vert }{1-r^{2}e^{-2\left\vert \mathbf{p}
\right\vert L}}\nonumber\\
&  \times\left\{  2\left(  1-r^{2}\right)  \left(  1-c\right)  ^{2}
r^{4}e^{-2\left\vert \mathbf{q}\right\vert L}e^{-2\left\vert \mathbf{p}
\right\vert L}\right.  \nonumber\\
&  +\left.  \left(  2r^{2}\left(  1-c\right)  ^{2}-2r\left(  1-c^{2}\right)
+4c\right)  r^{2}\left(  e^{-2\left\vert \mathbf{q}\right\vert L}
+e^{-2\left\vert \mathbf{p}\right\vert L}\right)  \right\}  \nonumber\\
c &  =\frac{\mathbf{q.p}}{\left\vert \mathbf{q}\right\vert \left\vert
\mathbf{p}\right\vert }\quad,\quad r\left[  i\xi\right]  =\frac{1-\epsilon
\left[  i\xi\right]  }{1+\epsilon\left[  i\xi\right]  }=-\frac{\omega
_{P}^{2}}{\omega_{P}^{2}+2\xi^{2}}
\end{align}
The denominators in $T$ describe the resonances of the Fabry-Perot cavity
formed by the two mirrors and $c$ is the cosine of the angle between the two
transverse wavevectors $\mathbf{q}$ and $\mathbf{p}$. We finally rewrite
(\ref{EPPplasmon}) under the general form (\ref{EPP}) with $G$ obtained
through a convolution on wavevectors and $\rho\left[  \mathbf{k}\right]  $
deduced as the ratio of $G\left[  \mathbf{k}\right]  $ to 
$G\left[  \mathbf{0}\right]  $
\begin{equation}
G\left[  \mathbf{k}\right]  =-\hbar A\int\frac{\mathrm{d}^{2}\mathbf{q}}
{4\pi^{2}}\int\limits_{0}^{\infty}\frac{\mathrm{d}\xi}{2\pi}\ T\left[
\mathbf{q},\mathbf{k-q},i\xi\right]  \quad,\quad\rho\left[  \mathbf{k}\right]
=\frac{G\left[  \mathbf{k}\right]  }{G\left[  \mathbf{0}\right]  }
\end{equation}

We have computed numerically the variation of $\rho$ versus the dimensionless
factor $\left\vert \mathbf{k}\right\vert L$ and shown the result as the dashed
line on Figure \ref{Rho}. The PFA sector still corresponds to $\left\vert
\mathbf{k}\right\vert L\lesssim1$. Outside this sector, the factor $\rho$ is
not only larger than unity, but also larger than in the limit of long
distances. The behaviour of $\rho$ for large wavevectors is found to obey 
the same law as in (\ref{LimLong}) with a larger value for the coefficient $\beta$
\begin{equation}
\rho\left[  \mathbf{k}\right]  \simeq\beta\left\vert \mathbf{k}\right\vert
L\quad,\quad\beta\simeq0.45 \quad,\quad \left\vert \mathbf{k}\right\vert
L\gtrsim1 \label{LimShort}
\end{equation}

\section{Discussion}

We discuss the significance of these results for theory-experiment comparison, 
and focus our attention on the relative effect of roughness measured by 
(\ref{FPS}) when the two conditions $R\gg L$ and $RL\gg\ell_{C}^{2}$ 
are met. If the stronger condition
$L\lesssim\ell_{C}$ is true, this effect is given by the PFA
\begin{equation}
\frac{\mathcal{F}_{\mathrm{PS}}-F_{\mathrm{PS}}}{F_{\mathrm{PS}}} \simeq \frac
{L^{2}E_{\mathrm{PP}}^{\prime\prime}}{2E_{\mathrm{PP}}}\frac{a^{2}}{L^{2}}
\label{FPSpfa}
\end{equation}
It is proportional to two dimensionless factors: the first one $\frac
{L^{2}E_{\mathrm{PP}}^{\prime\prime}}{2E_{\mathrm{PP}}}$ varies from 6 at long
distances (where $E_{\mathrm{PP}} \propto 1/L^{3}$) to 3 at short
distances (where $E_{\mathrm{PP}} \propto 1/L^{2}$) and the second one
is the square $a^{2}/L^{2}$ of the ratio of roughness amplitude to
cavity length. This expression is equivalent to the procedure
which has been used for analyzing recent experiments
\cite{Harris00,Klimchitskaya99} and led to a relative effect equal to a
fraction of a percent. It is now clear that this procedure underestimates the
effect as soon as the roughness spectrum contains wavevectors $\left\vert
\mathbf{k}\right\vert L\gtrsim1$. This case is not excluded by a preliminary 
inspection of available images of the rough plates used in the experiments.

We then have to use the expression (\ref{FPS}) which includes a third factor
$\overline{\rho}$ representing the modification of the
correction with respect to the PFA  
\begin{equation}
\frac{\mathcal{F}_{\mathrm{PS}}-F_{\mathrm{PS}}}{F_{\mathrm{PS}}}
=\frac{L^{2}E_{\mathrm{PP}}^{\prime\prime}}{2E_{\mathrm{PP}}}
\frac{a^{2}}{L^{2}} \overline{\rho} \quad,\quad 
\overline{\rho} = \int\frac{\mathrm{d}^{2}\mathbf{k}}{4\pi^{2}}
\ \rho\left[  \mathbf{k}\right]  \ \frac{\sigma\left[  \mathbf{k}\right]}{a^{2}}
\label{DeltaF}
\end{equation}
$\overline{\rho}$ is the mean value of the roughness sensitivity 
$\rho\left[  \mathbf{k}\right]  $ in the distribution given by the normalized 
roughness spectrum $\sigma\left[\mathbf{k}\right] /a^{2}$. 
To give an illustration, it takes the simple form
$\overline{\rho}\simeq\beta\sqrt{\pi} L/\ell_{C}$ for a
Gaussian roughness spectrum (\ref{Gaussian}) having an important weight
outside the PFA sector. It follows that the roughness correction has
completely different scaling behaviours with respect to $L$ inside and outside
the PFA sector: (\ref{FPSpfa}) scales as $\frac{a^{2}}{L^{2}}$ when
$L\lesssim\ell_{C}$ whereas (\ref{DeltaF}) scales as $\beta\sqrt{\pi
}\frac{a^{2}}{L\ell_{C}}$ when $L\gtrsim\ell_{C}$ (see
similar discussions in \cite{Emig01,Novikov92}).

We stress again that the proximity force approximation underestimates the
sensitivity of the Casimir effect to roughness. In order to ensure that the
informations deduced from accurate theory/experiments comparisons are not
biased by this approximation, it is certainly necessary to evaluate the factor
$\overline{\rho}$. This requires the measurement of the normalized roughness
spectrum $\sigma\left[  \mathbf{k}\right] /a^{2}$ for the plates used
in the experiments as well as the computation of the normalized roughness
sensitivity function $\rho\left[  \mathbf{k}\right]  $. Here, this function
$\rho\left[  \mathbf{k}\right]  $ has been obtained in the two limiting cases
of long or short distances by developing previous calculations. It has been found
to be different in the two cases, with a deviation from PFA
more pronounced in the short distance limit where the Casimir
effect is also more sensitive to roughness. This means that a more general
evaluation of $\rho\left[  \mathbf{k}\right]  $, covering the experimentally
explored range of values of $L/\lambda_{P}$, is required for
a reliable estimation of the effect of roughness.

\acknowledgments CG acknowledges CAPES and COFECUB for having sponsored his
stay in Rio de Janeiro. PMN acknowledges financial support by ENS and PRONEX
which made possible his stay in Paris. PMN also acknowledges partial financial
support by the Instituto do Milenio de Informacao Quantica and CNPq.

\end{document}